\documentclass[conference,a4paper]{IEEEtran}

\usepackage{amsmath}
\usepackage{amssymb}
\usepackage{latexsym}
\usepackage{array,arydshln}
\usepackage{multirow}
\usepackage{graphicx}
\usepackage{float}
\usepackage{bm}
\usepackage{breqn}
\usepackage{ctable}
\usepackage{color}
\usepackage{url}
\usepackage{cite}

\newcommand{\otoprule}{\midrule[\heavyrulewidth]}
\newcommand{\bs}[1]{\ensuremath{\boldsymbol{#1}}}

\newcommand{\beq}{\begin{equation}}
\newcommand{\enq}{\end{equation}}
\newcommand{\beal}{\begin{align*}}
\newcommand{\enal}{\end{align*}}

\IEEEoverridecommandlockouts

\begin{document}

\title{A Unified Ensemble of Concatenated \\Convolutional Codes}


\author{
\IEEEauthorblockN{Saeedeh Moloudi$^\dag$, Michael Lentmaier$^\dag$, and Alexandre Graell i Amat$^\ddag$}
\IEEEauthorblockA{$\dag$Department of Electrical and Information
  Technology, Lund University, Lund, Sweden \\
  $\ddag$Department of Signals and Systems, Chalmers University of Technology, Gothenburg, Sweden\\
              \{saeedeh.moloudi,michael.lentmaier\}@eit.lth.se, alexandre.graell@chalmers.se}\\
              \thanks{This work was supported in part by the Swedish Research Council (VR) under grant \#621-2013-5477.}\vspace*{-1cm}
}


\maketitle

\begin{abstract}
We introduce a unified ensemble for turbo-like codes (TCs) that contains the four main classes of TCs: parallel concatenated codes, serially concatenated codes, hybrid concatenated codes, and braided convolutional codes.
    We show that for each of the original classes of TCs, it is possible to find an equivalent ensemble by proper selection of the design parameters in the unified ensemble. 
    We also derive the density evolution (DE) equations for this ensemble over the binary erasure channel.
    The thresholds obtained from the DE indicate that the TC ensembles from the unified ensemble have similar asymptotic behavior to the original TC ensembles.

\end{abstract}

\IEEEpeerreviewmaketitle

\section{Introduction}

Over the last few years, research on low-density parity-check (LDPC) convolutional codes \cite{JimenezLDPCCC}, also known as spatially coupled LDPC (SC-LDPC) codes \cite{Kudekar_ThresholdSaturation}, has become very popular.
It is proved that for these codes, the belief propagation (BP) decoder can achieve the threshold of the maximum-a-posteriori (MAP) decoder \cite{Kudekar_ThresholdSaturation,Yedla2014}.
This remarkable phenomenon is known as threshold saturation.
Spatial coupling is a general concept that is not limited to LDPC codes. 
Recently, spatially coupled turbo-like codes (SC-TCs) were introduced in \cite{Moloudi_SCTurbo, Moloudi_SPCOM14, JournalMLD}.
In these works, the spatial coupling of the four main classes of TCs was considered.
These included parallel concatenated codes (PCCs) \cite{BerrouTC}, serially concatenated codes (SCCs) \cite{Benedetto98Serial}, braided convolutional codes (BCCs)\cite{ZhangBCC}, and hybrid concatenated codes (HCCs) \cite{DivsalarHCC,KollerHCC}.
The density evolution (DE) analysis performed in \cite{Moloudi_SCTurbo, Moloudi_SPCOM14, JournalMLD} suggests that SC-TCs have an excellent asymptotic behavior and for them, threshold saturation occurs.
This gives a new perspective in designing a concatenated coding scheme: optimizing the uncoupled ensembles for achieving the best BP threshold may not necessarily lead to the best overall performance.

TCs are adopted in many communication standards.
Each class of TCs exhibits a unique asymptotic behavior.
While some classes---such as PCCs---yield good BP thresholds some others---such as SCCs and BCCs---have excellent MAP thresholds.
As shown in \cite{Moloudi_SCTurbo, Moloudi_SPCOM14, JournalMLD}, spatial coupling improves the thresholds of the TC classes. 

So far, the different classes of TCs have been considered separately.
A unified ensemble which contains all main TC ensembles can unify the frameworks for analysis, and clarify the connections between the TC classes.
In fact, this ensemble can lead to a better understanding of the similarities and differences between various TC classes and the possible trade-offs in the code design.
In \cite{AlexUnifying}, the authors introduced an ensemble which unifies PCCs and SCCs. This ensemble is based on concatenations of several component encoders and does not cover the BCC and HCC ensembles.

In this paper, we introduce an ensemble of concatenated convolutional codes that encompasses all the above-mentioned four major classes of TCs.
For simplicity, we only use a single rate-1 component code.
In other words, the different trellises are combined to a single self-concatenated trellis.
Probably, the most famous class of self-concatenated convolutional codes are repeat accumulate (RA) codes, first introduced in \cite{TurboLike98}.
This class of codes is covered by the proposed ensemble
if the component code in the equivalent PCC ensemble is set as an accumulator.
In order to find a self-concatenated equivalent for the other classes of TCs, some feedback path has to be introduced in the encoder structure. 
 The differences between the various original TC ensembles are then reflected in the permutation structure and the amount of feedback in the unified ensemble. 

We also derive the exact density evolution equations for the binary erasure channel (BEC).
Using these equations, we compute the BP thresholds of the corresponding classes of TCs and we show that the obtained thresholds are very close to the thresholds of the original ensembles.
 
\section{Steps toward the Self-Concatenated Ensemble}
In this section, for each class of TCs, we separately describe how to reduce the number of component codes in order to obtain the equivalent self-concatenated ensemble.  
\subsection{Parallel Concatenated Codes}
\begin{figure}[t]
	\centering
	\includegraphics[width=0.85\linewidth]{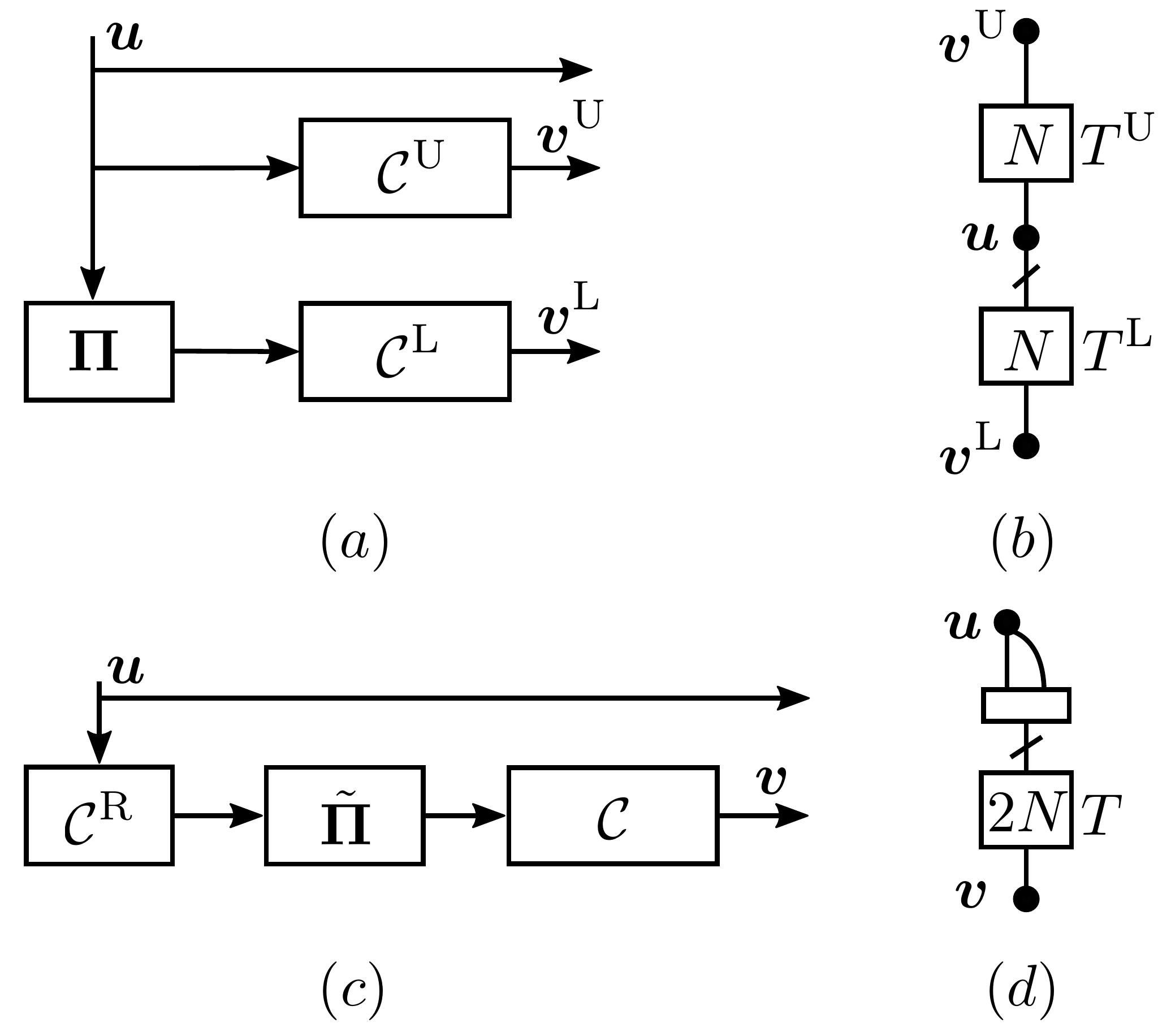}
	\caption{(a) Block diagram of a PCC, (b) Compact graph of a PCC, (c) Block diagram of a self-concatenated PCC, and (d) Compact graph of a self-concatenated PCC }
	\label{PCC}
\end{figure}
Fig.~\ref{PCC}(a) and (b) show the encoder block diagram and compact graph representation \cite{JournalMLD} of a PCC, respectively.
The considered PCC ensemble is built of two identical rate-1 component encoders, called upper and lower encoders, and shown by $\mathcal{C}^{\text{U}}$ and $\mathcal{C}^{\text{L}}$, respectively.
The information sequence $\bs{u}$ is connected to $\mathcal{C}^{\text{U}}$ to produce the parity sequence $\bs{v}^{\text{U}}$.
Likewise, a reordered copy of $\bs{u}$ is connected to $\mathcal{C}^{\text{L}}$ to produce the parity sequence $\bs{v}^{\text{L}}$. 
The output of the PCC encoder is the tuple $(\bs{u},\bs{v}^{\text{U}},\bs{v}^{\text{L}})$.  
In the compact graph representation (see Fig.~\ref{PCC}(b)), the trellises corresponding to  $\mathcal{C}^{\text{U}}$ and  $\mathcal{C}^{\text{L}}$, are depicted by squares (factor nodes) and denoted by $T^{\text{U}}$ and $T^{\text{L}}$, respectively. 
These factor nodes are labeled with the length of the corresponding trellises.
Each of the sequences $\bs{u}$, $\bs{v}^{\text{U}}$, and $\bs{v}^{\text{L}}$ is represented by a black circle, called variable node. The permutation $\bs{\Pi}$ in the block diagram is replaced in Fig.~\ref{PCC}(b) by a line that crosses the edge between $\bs{u}$ and $\bs{v}^{\text{L}}$.

Fig.~\ref{PCC}(c) and (d) show respectively the encoder block diagram and compact graph representation of the self-concatenated coding ensemble corresponding to a PCC.  
In this ensemble, the two component encoders of the PCC ensemble are replaced by a repetition encoder $\mathcal{C}^{\text{R}}$ (with repeating factor $2$) followed by a rate-1 convolutional encoder $\mathcal{C}$. 
The information sequence $\bs{u}$ is connected to  $\mathcal{C}^{\text{R}}$ to produce the sequence $(\bs{u},\bs{u})$. The resulting sequence is reordered by the permutation $\tilde{\bs{\Pi}}$ and used as input to $\mathcal{C}$.
The parity sequence $\bs{v}$ has length $2N$ and corresponds to both $\bs{v}^{\text{U}}$ and $\bs{v}^{\text{L}}$ in the original ensemble.
Note that, by replacing $\mathcal{C}$ by an accumulator in Fig.~\ref{PCC}(c), a RA code can be obtained.
However, the ensemble in Fig.~\ref{PCC}(c) is more general as $\mathcal{C}$ can be any convolutional encoder.

In the compact graph representation (see Fig.~\ref{PCC}(d)), the repetition of the information sequence is shown by increasing the degree of the corresponding variable node. As it is shown in the figure, the sequence $\bs{u}$ and its repetition are multiplexed to produce the sequence $(\bs{u},\bs{u})$.
The multiplexer is represented by a rectangle.
The resulting sequence is connected to trellis $T$ to produce the parity sequence $\bs{v}$. 
Note that the length of the trellis in the self-concatenated ensemble is twice of the length of $T^{\text{U}}$ and $T^{\text{L}}$ in the original ensemble.

In this paper we consider block-wise multiplexers.
By selecting 
\begin{equation}
\tilde{\bs{\Pi}}=\left[\begin{array}{cc}\bs{I}&0\\0&\bs{\Pi}\end{array}\right],
\label{Pi}
\end{equation}
 where $\bs{I}$ is the $N\times N$ identity matrix, the self-concatenated ensemble is equivalent to the original ensemble.

\subsection{Serially Concatenated Codes}
\begin{figure}[t]
	\centering
	\includegraphics[width=1\linewidth]{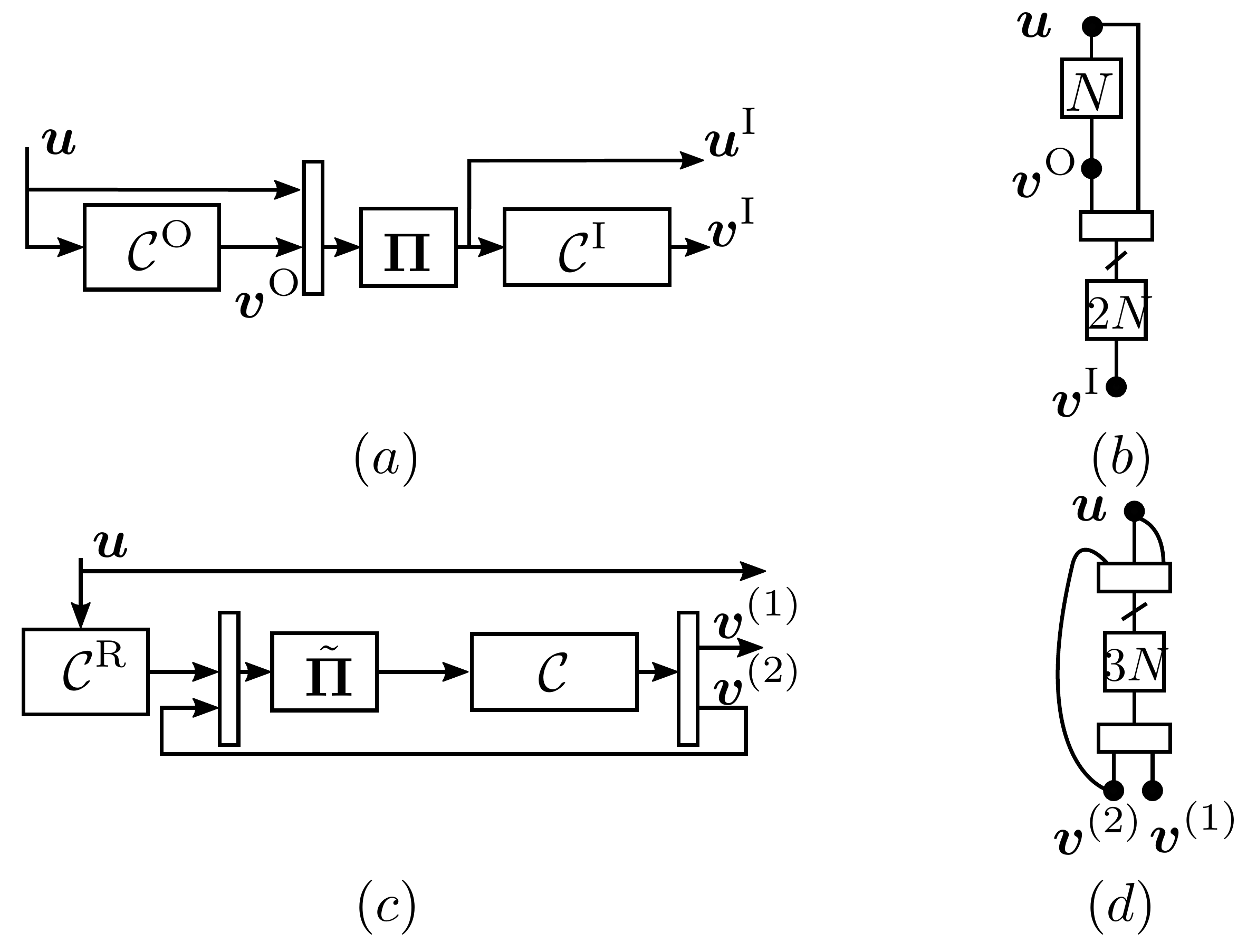}
	\caption{(a) Block diagram of a SCC, (b) Compact graph of a SCC, (c) Block diagram of a self-concatenated SCC, and (d) Compact graph of a self-concatenated SCC}
	\label{SCC}
\end{figure}
The encoder block diagram and the compact graph representation of the SCC ensemble are shown in 
Fig.~\ref{SCC}(a) and (b), respectively.
This ensemble is built of two identical rate-1 component encoders called outer and inner encoders and shown by $\mathcal{C}^{\text{O}}$ and $\mathcal{C}^{\text{I}}$, respectively.
The length-$N$ information sequence, $\bs{u}$, is connected to $\mathcal{C}^{\text{O}}$ to produce the parity sequence $\bs{v}^{\text{O}}$. Then, the sequences $\bs{u}$ and $\bs{v}^{\text{O}}$ are multiplexed and reordered. The resulting sequence is used as input for $\mathcal{C}^{\text{I}}$ to produce the parity sequence $\bs{v}^{\text{I}}$.

The encoder block diagram and compact graph representation of the equivalent self-concatenated ensemble are shown in Fig.~\ref{SCC}(c) and (d), respectively.
In this ensemble, the trellises of the outer and inner encoders are combined to make a  trellis with length $3N$. 
Similarly to the self-concatenated ensemble for PCCs, $\bs{u}$ is connected to $\mathcal{C}^{\text{R}}$ with repetition factor $2$ to produce the sequence $\bs{\tilde{u}}=(\bs{u},\bs{u})$.
In the original ensemble $\bs{v}^{\text{O}}$ is used as part of the input to $\mathcal{C}^{\text{I}}$. 
To satisfy this condition with only one component encoder, the overall parity sequence of the self-concatenated ensemble, $\bs{v}$, is divided into two sequences  $\bs{v}^{(1)}$ and $\bs{v}^{(2)}$, of length $2N$ and $N$, respectively. 
Then, $\bs{v}^{(2)}$ is used as a part of the input sequence through a feedback path.
The feedback path connects $\bs{v}^{(2)}$ to a multiplexer.
Then, this sequence is multiplexed with sequence $\bs{\tilde{u}}$.
The resulting sequence is reordered by $\tilde{\bs{\Pi}}$ and used as input to $\mathcal{C}$. Note that sequences $\bs{v}^{(1)}$ and $\bs{v}^{(2)}$ correspond to $\bs{v}^{\text{O}}$ and $\bs{v}^{\text{I}}$ in the original ensemble, respectively.

We remark that, in general, the encoder of the self-concatenated ensemble is not causal. However, this problem can be solved by proper selection of $\tilde{\bs{\Pi}}$ or by spatial coupling.
By selecting $\tilde{\bs{\Pi}}$ as in \eqref{Pi} the corresponding trellis is split into two parts. The information sequence $\bs{u}$ is connected to the first part to produce $\bs{v}^{(2)}$. Then, a reordered copy of the sequence $(\bs{u},\bs{v}^{(2)})$ is connected to the second part of the trellis to produce $\bs{v}^{(1)}$.
By spatial coupling, the feedback path can be fed forward to the corresponding multiplexer in the next time slots.
 
\subsection{Hybrid Concatenated Codes}
\begin{figure}[t]
	\centering
	\includegraphics[width=0.95\linewidth]{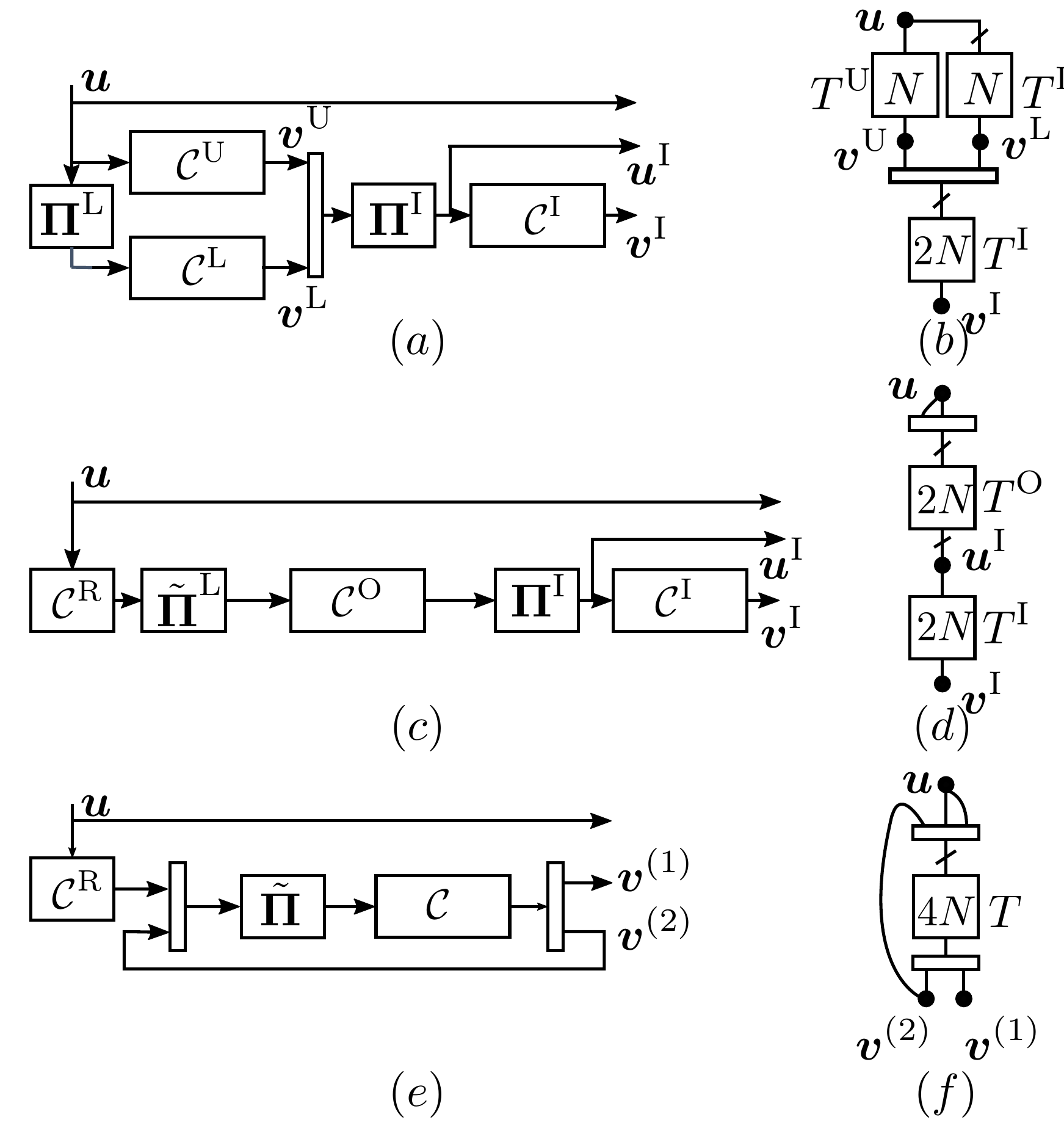}
	\caption{(a) Block diagram of a HCC, (b) Compact graph a HCC, (c) Block diagram of a self-concatenated HCC (step 1), (d) Compact graph of a self-concatenated HCC, (e) Block diagram of a self-concatenated HCC (step 1), and (f) Compact graph of a self-concatenated HCC}
	\label{HCC}
\end{figure}
Fig.~\ref{HCC}(a) shows the encoder block diagram of an HCC ensemble built from three identical rate-1 component encoders. 
The corresponding compact graph representation is also shown in Fig.~\ref{HCC}(b).
The considered HCC ensemble is a serial concatenation of a parallel ensemble with an inner encoder. 
The information sequence $\bs{u}$ and a reordered copy of it are fed to two encoders, referred to as upper and lower encoders, and denoted by $\mathcal{C}^{\text{U}}$ and $\mathcal{C}^{\text{L}}$, to produce parity sequences $\bs{v}^{\text{U}}$ and $\bs{v}^{\text{L}}$, respectively.
Then, $\bs{v}^{\text{U}}$ and $\bs{v}^{\text{L}}$ are multiplexed and reordered. The resulting sequence is used as an input to the inner encoder $\mathcal{C}^{\text{I}}$. 

The corresponding self-concatenated ensemble can be obtained in two steps. 
First, as it is shown in Fig.~\ref{HCC}(c)(d), the upper and lower trellises can be unified into a single trellis with length $2N$ by the method described for PCCs.
Then, the resulting trellis can be connected to the inner trellis using the method described for SCCs. 

The self-concatenated ensemble for HCCs is shown in Fig.~\ref{HCC}(e)(f).
In this ensemble, the overall trellis has length $4N$.
The parity sequence $\bs{v}$ is divided into two equal-size sequences $\bs{v}^{(1)}$ and $\bs{v}^{(2)}$ of length $2N$ . 
Then, $\bs{v}^{(2)}$ is multiplexed with sequence $(\bs{u},\bs{u})$ generated by a repetition encoder $\mathcal{C}^{\text{R}}$.
The resulting sequence is reordered and used as an input to a rate-$1$ convolutional encoder $\mathcal{C}$.
Note that $\bs{v}^{(1)}$ and $\bs{v}^{(2)}$ correspond to $(\bs{v}^{\text{U}}, \bs{v}^{\text{L}})$  and  $\bs{v}^{\text{I}}$ of the original ensemble. By selecting
\begin{equation}
\tilde{\bs{\Pi}}=\left[\begin{array}{ccc}\bs{I}&0&0\\0&\bs{\Pi}^{\text{L}}&0\\0&0&\bs{\Pi}^{\text{I}}\end{array}\right],
\label{Pi2}
\end{equation}
the self-concatenated ensemble is equivalent to the original ensemble.

\subsection{Braided Convolutional Codes}
\begin{figure}[t]
    \centering
	\includegraphics[width=0.95\linewidth]{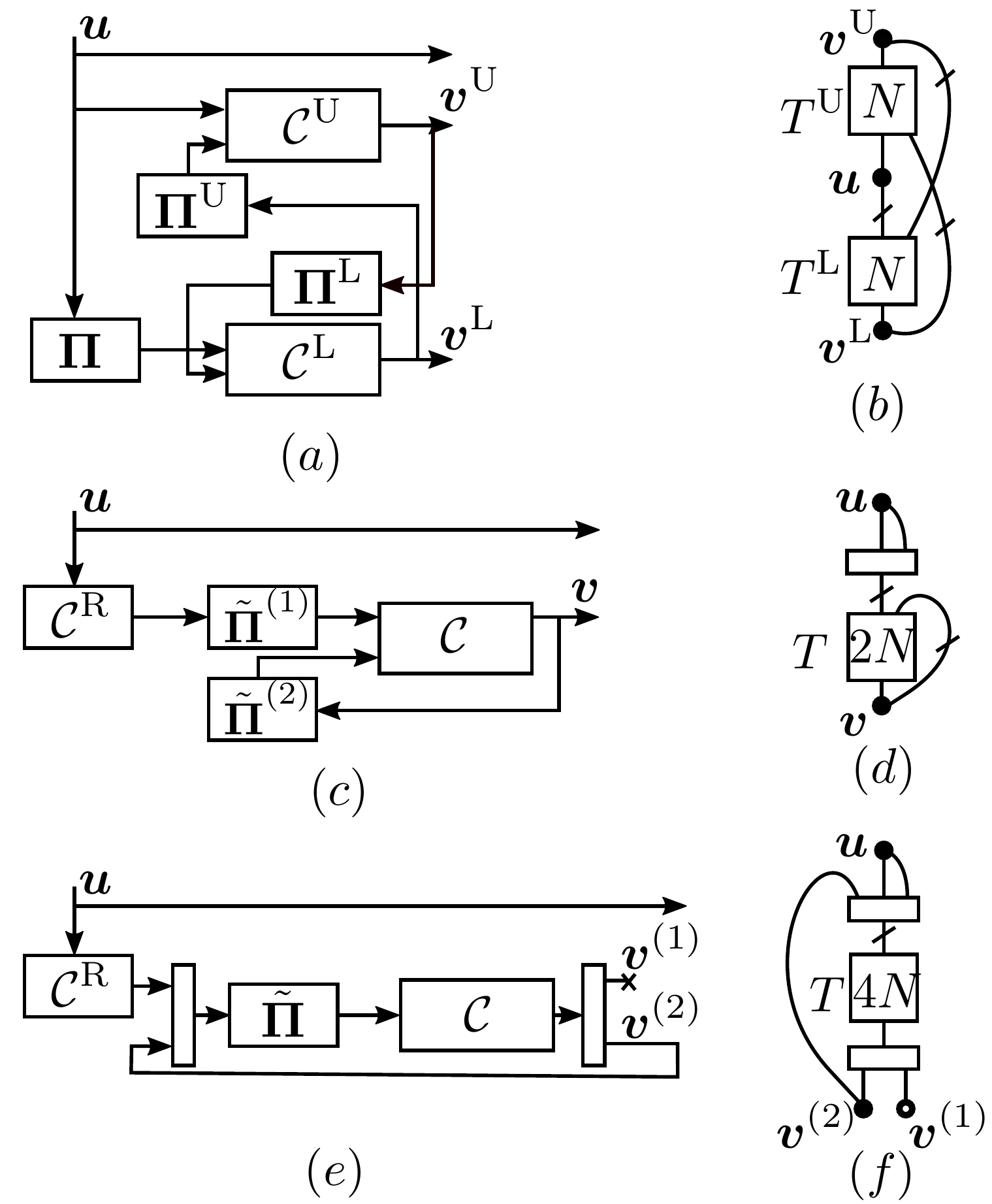}
	\caption{(a) Block diagram of a BCC, (b) Compact graph a BCC, (c) Block diagram of a self-concatenated BCC (step 1), (d) Compact graph of a self-concatenated BCC, (e) Block diagram of a self-concatenated BCC (step 1), and (f) Compact graph of a self-concatenated BCC}
	\label{BCC}
\end{figure}
Fig.~\ref{BCC}(a) shows the encoder block diagram of a BCC ensemble. 
This ensemble is similar to the PCC ensemble but the BCC ensemble is built of two rate-$2$ convolutional encoders and the parity sequence of each encoder is fed back to the input of the other encoder.
Similarly to PCCs, the component encoders are denoted by $\mathcal{C}^{\text{U}}$ and $\mathcal{C}^{\text{L}}$ and called upper and lower encoders, respectively. 
The compact graph representation of the ensemble is shown in Fig.~\ref{BCC}(b). 
The information sequence $\bs{u}$ and a reordered copy of $\bs{v}^{\text{L}}$ are used as the first and second input of $\mathcal{C}^{\text{U}}$, respectively to produce the parity sequence $\bs{v}^{\text{U}}$.
Likewise, a reordered copy of  $\bs{u}$ and a reordered copy of $\bs{v}^{\text{U}}$ are used as the first and second input of $\mathcal{C}^{\text{L}}$ respectively, to produce $\bs{v}^{\text{L}}$.

In order to obtain the self-concatenated ensemble for BCCs, we can use the method described for PCCs.
Fig.~\ref{BCC}(c) and (d) show the encoder block diagram and compact graph representation of the corresponding self-concatenated ensemble.
The two component encoders, with $N$ trellis sections, in the original ensembles are combined to a component encoder with length-$2N$ trellis. 
The sequence $\bs{u}$ is connected to a repetition encoder $\mathcal{C}^{\text{R}}$ to produce the sequence $(\bs{u},\bs{u})$.
The resulting sequence is reordered by permutation $\tilde{\bs{\Pi}}^{(1)}$ and used as the first input of a rate-2 convolutional encoder $\mathcal{C}$.
The second input of the encoder is a copy of the parity sequence $\bs{v}$ that is reordered by the permutation $\tilde{\bs{\Pi}}^{(2)}$.
By selecting the permutations as
\[
\tilde{\bs{\Pi}}^{(1)}=\left[\begin{array}{cc}\bs{I}&0\\0&\bs{\Pi}\end{array}\right],
\]
\[
\tilde{\bs{\Pi}}^{(2)}=\left[\begin{array}{cc}0&\bs{\Pi}^{\text{U}}\\\bs{\Pi}^{\text{L}}&0\end{array}\right],
\]
 the encoders in Fig.~\ref{BCC}(a) and (c) are equivalent.

The encoder ensembles in Fig.~\ref{BCC}(a) and (c) are not causal. 
For the original ensemble of BCCs introduced in \cite{ZhangBCC}, this problem was solved by spatial coupling.
In the block-wise BCC ensemble, the parity sequences $\bs{v}^{\text{U}}$ and $\bs{v}^{\text{L}}$, (or $\bs{v}$ in the self-concatenated ensemble), are connected to the inputs of the corresponding encoders after passing delay blocks. This makes the encoder causal. 

It is also possible to find a self-concatenated ensemble for BCCs that is very close to those for the other TC classes.
We can replace the rate-$2$ component encoder in the self-concatenated ensemble by a rate-$1$ encoder for that half of its output sequence is punctured.
The encoder block diagram for this ensemble is shown in Fig.~\ref{BCC}(e) and its corresponding compact graph representation is depicted in Fig.~\ref{BCC}(f).
As it is shown in these figures, the parity sequence is divided into two parts $\bs{v}^{(1)}$ and $\bs{v}^{(2)}$.
Sequence $\bs{v}^{(1)}$ is fully punctured and
$\bs{v}^{(2)}$ is multiplexed with the sequence $(\bs{u},\bs{u})$ at the output of the repetition encoder $\mathcal{C}^{\text{R}}$.
 The resulting sequence is reordered and fed to the convolutional encoder $\mathcal{C}$ with corresponding trellis of length $4N$.

\section{The Unified Ensemble}\label{UnifiedEnsemble}
\begin{figure}[t]
	\centering
	\includegraphics[width=0.95\linewidth]{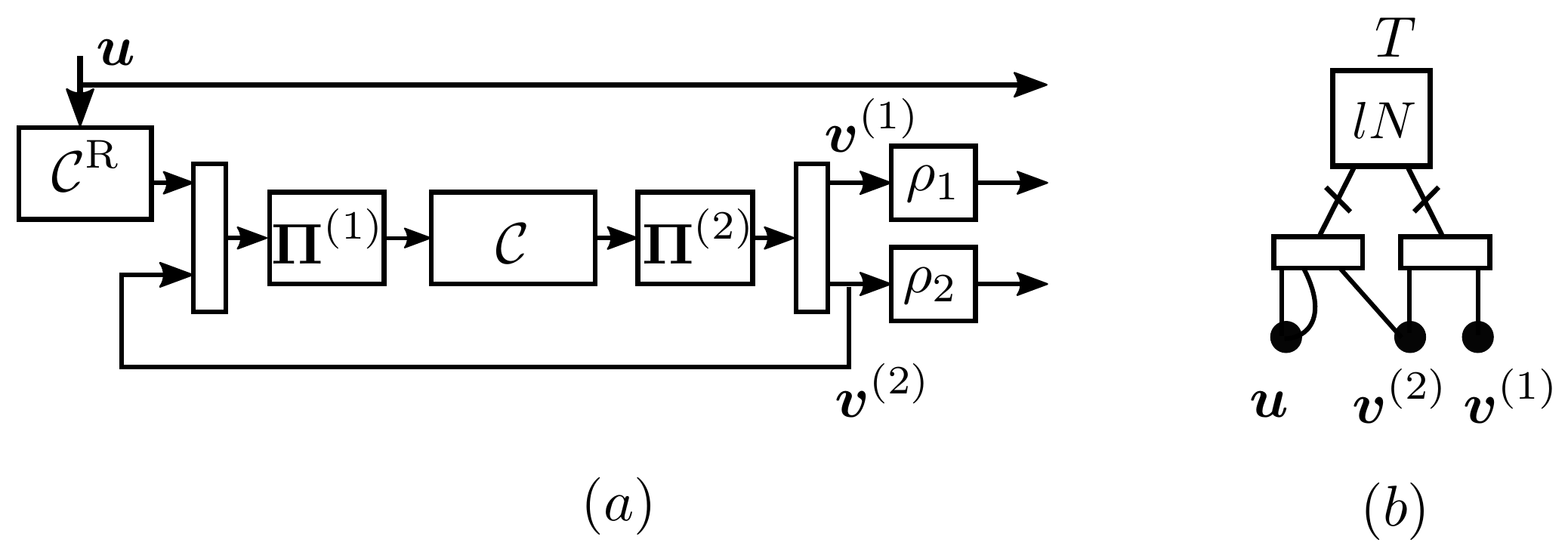}
	\caption{(a) Block diagram of encoder  (b) Compact graph representation of the unified ensemble  }
	\label{Block}
\end{figure}
Comparing the obtained self-concatenated ensembles introduced in the previous section for the considered classes of TCs, we note similarities between them (see Fig.~\ref{PCC}-Fig.~\ref{BCC}). 
Based on these similarities, we develop a unified ensemble.
The block diagram of the encoder and the compact graph representation of this ensemble is shown in Fig.~\ref{Block}.
In this ensemble, the component convolutional encoder is a rate-$1$ convolutional encoder. The information sequence $\bs{u}$, of length $N$, is connected to a repetition encoder $\mathcal{C}^{\text{R}}$ with repetition factor $2$ to produce the sequence $\tilde{\bs{u}}=(\bs{u},\bs{u})$.
This sequence is multiplexed with $\bs{v}^{(2)}$ which is a part of the overall parity sequence $\bs{v}$. 
Then, the resulting sequence is reordered by the permutation $\bs{\Pi}^{(1)}$, of size $lN$, and fed to the convolutional encoder $\mathcal{C}$.
Parameter $l$ is a design parameter which can be tuned to yield a specific TC class (PCC, SCC, BCC, or HCC).
The values of $l$ for the different classes of TCs are presented in Table \ref{GEnsemble}.

The sequence $\bs{v}$ is divided into two sequences $\bs{v}^{(1)}$ and $\bs{v}^{(2)}$ of length $l_1N$ and $l_2N$, respectively. In order to guarantee random division of $\bs{v}$, first, this sequence is reordered by a permutation. Then, it is divided into the two sequences $\bs{v}^{(1)}$ and $\bs{v}^{(2)}$.
The values of $l_1$ and $l_2$ for each class of TCs, are provided in Table \ref{GEnsemble}.
Note that $l=l_1+l_2$.  

Parameters $\rho_1$ and $\rho_2$ in Table \ref{GEnsemble} are the permeability rates for sequences $\bs{v}^{(1)}$ and $\bs{v}^{(2)}$, respectively, giving the fraction of surviving bits of $\bs{v}^{(1)}$ and $\bs{v}^{(2)}$ after puncturing. 
For example, to obtain the equivalent ensemble for BCCs, $\rho_1=0$, as $\bs{v}^{(1)}$ is fully punctured.
We remark that by selecting $\rho_1$ and $\rho_2$ properly, the obtained ensemble covers a family of rate compatible TCs.
\begin{table}[t]	
	\caption{Parameters of PCCs, SCCs, BCCs and HCCs }
	\begin{center}
		\begin{tabular}{lccccccc}
			\toprule
			Ensemble&$R$&$\rho_1$&$\rho_2$&$l$&$l_1$&$l_2$&$l_2/l_1$\\
			\otoprule
			PCC&1/3&1&-&2&2&0&0\\[0.5mm]\midrule
			SCC& 1/4&1&1&3&2&1&1/2\\[0.5mm]\midrule
			BCC&1/3&0&1&4&2&2&1\\[0.5mm]\midrule
			HCC&1/5&1&1&4&2&2&1\\[0.5mm] \bottomrule
		\end{tabular}
		\label{TableUnified}
	\end{center}
	\label{GEnsemble} 
	\vspace{-3.5ex}
\end{table}

An inspection of Fig.~\ref{Block}(b) reveals that the compact graph representation of the unified ensemble is very close to the protograph of an irregular LDPC code.
The factor node is a trellis with length $lN$, where its degree is fixed to two.
The variable nodes are classified into three groups as follows: $\bs{u}$ is an information variable node with degree $2$, $\bs{v}^{(1)}$ is a parity variable node with degree $1$, and $\bs{v}^{(2)}$ is a parity variable node with degree $2$.
The length of theses variable nodes are not equal.
Considering length $N$ for $\bs{u}$, $\bs{v}^{(1)}$ and $\bs{v}^{(2)}$ have length $l_1N$ and $l_2N=2N$, respectively. 
According to Table \ref{TableUnified}, different ensembles of TCs can be obtained by changing the ratio $l_2/l_1$ which is the proportion of degree-$1$ and degree-$2$ parity variable nodes.   
This is very close to defining the variable node degree distribution for LDPC codes.  

\subsection{Density Evolution}
Considering transmission over a BEC with channel parameter $\varepsilon$, we can analyze the asymptotic behavior of the unified ensemble by tracking the evolution of the erasure probability with the number of iterations of the decoding procedure. This evolution can be shown as a set of equations called DE equations, and for the BEC, it is possible to derive an exact expression for them.
In the $i$th iteration, the extrinsic erasure probabilities from factor node $T$ toward variable nodes are denoted by $x_1^{(i)}$ and $x_2^{(i)}$, respectively, for the first and second edge connected to it.
Then, the DE equations can be written as, 
\begin{align}
x_1^{(i+1)}=f_1(y_1^{(i)},y_2^{(i)}),\\
x_2^{(i+1)}=f_2(y_1^{(i)},y_2^{(i)})
\end{align}
where
\begin{align}
&y_1^{(i)}=\frac{2\varepsilon x_1^{(i)}+l_2(\rho_2 \varepsilon x_2^{(i)}+(1-\rho_2)x_2^{(i)})}{2+l_2},\label{DE3}
\\
&y_2^{(i)}=\frac{l_2(\rho_2 \varepsilon x_1^{(i)}+(1-\rho_2)x_1^{(i)})+l_1(\rho_1 \varepsilon+(1-\rho_1))}{2+l_2}.\label{DE4}
\end{align}

Here, $f_1$ and $f_2$ are the transfer functions of $T$ for the systematic and parity bits, respectively. The a-posteriori erasure probability of bits in the information sequence $\bs{u}$ at the $i$th iteration is,
\[
p_a^{(i)}=\varepsilon \cdot (x_1^{(i)})^2.
\] 
	
\begin{table}[t]	
	\caption{Thresholds of PCCs, SCCs, BCCs and HCCs }
	\begin{center}
		\begin{tabular}{lccccc}
			\toprule
			Ensemble&$R$&$\varepsilon_{\text{BP}}$ &$\varepsilon_{\text{BP}}^{\text{U}}$& $\varepsilon_{\text{MAP}}$&$\varepsilon_{\text{MAP}}^{\text{U}}$ \\
			\otoprule
			PCC&1/3&0.6428&0.6428&0.6553&0.6552\\[0.5mm]\midrule
			SCC& 1/4&0.6895&0.6863&0.7481&0.7482\\[0.5mm]\midrule
			BCC&1/3&0.5541&0.5603&0.6653&0.6646\\[0.5mm]\midrule
			HCC&1/5&0.7261&0.6997&0.7995&0.7994\\[0.5mm]\bottomrule
		\end{tabular}
		\label{TableTH}
	\end{center}
	\label{BPThresholds} 
	\vspace{-3.5ex}
\end{table}
The decoding thresholds obtained by DE are reported in Table \ref{TableTH}.  
The table shows the BP threshold $\varepsilon_{\text{BP}}$ and the MAP threshold $\varepsilon_{\text{MAP}}$ of the original ensembles. 
To obtain $\varepsilon_{\text{BP}}$ and $\varepsilon_{\text{MAP}}$, we used the corresponding DE equations and the area theorem, respectively.
We also report in the table the BP threshold $\varepsilon_{\text{BP}}^{\text{U}}$ and the MAP threshold $\varepsilon_{\text{MAP}}^{\text{U}}$ of the proposed equivalent ensembles.
From the results in the table, it can be seen that in the PCC case there is a good match between the thresholds of the original ensemble and the corresponding values of the equivalent ensemble. For the other cases, $\varepsilon_{\text{BP}}$ and $\varepsilon_{\text{BP}}^{\text{U}}$ are similar. 
However, there is a small gap between these thresholds.
This gap can be explained as follows.
In the DE analysis of the unified ensemble, we consider that the permutations are chosen randomly.
Therefore, in equations \eqref{DE3} and \eqref{DE4}, we average over all possible cases. 
However, the original TC ensembles are more structured and, in consequence, except for BCCs, $\varepsilon_{\text{BP}}$ is larger than $\varepsilon_{\text{BP}}^{\text{U}}$.
For BCCs, replacing the rate-2 component encoder by a rate-1 component encoder with puncturing, is another reason for observing the gap between $\varepsilon_{\text{BP}}$ and $\varepsilon_{\text{BP}}^{\text{U}}$.
We also computed the thresholds for the self-concatenated ensemble in Fig.~\ref{BCC}(c). The obtained BP and MAP thresholds for this ensemble are identical to those of the original BCC ensemble.  

The results in Table \ref{TableTH} demonstrate that thresholds similar to those of the original TC classes can be obtained by changing the design parameters in the unified ensemble.
\section{Conclusions}
In this paper, a unified ensemble for various classes of TCs is introduced.
This ensemble is based on a single trellis with self-concatenation.
We introduced two elementary steps to find the self-concatenated equivalent of PCCs and SCCs.
We also used these elementary steps to find the self-concatenated HCCs and BCCs.
These elementary steps can also be applied to more general concatenations.

Then, by considering the similarities between the self-concatenated ensembles for different TC classes, we found a unified ensemble. 
 By changing the proportion of degree-$1$, and degree-$2$ variable nodes in the graph or puncturing a part of the parity sequence, the original TC ensembles can be obtained. 
 The compact graph representation of our ensemble establishes a bridge between TCs and protograph based LDPC codes, where the check nodes are replaced by trellis constraints.

We believe that the unified ensemble may help in better understanding the connections between concatenated code ensembles and LDPC code ensembles.


\end{document}